\documentclass[aps,pra,twocolumn,showpacs]{revtex4}
\usepackage{amsmath, amssymb}
\usepackage{color, graphicx}

\begin{document}

\title{True photo-counting statistics of multiple on/off detectors}
\author{J. Sperling} \email{jan.sperling2@uni-rostock.de}\affiliation{Arbeitsgruppe Quantenoptik, Institut f\"ur Physik, Universit\"at Rostock, D-18051 Rostock, Germany}
\author{W. Vogel}\affiliation{Arbeitsgruppe Quantenoptik, Institut f\"ur Physik, Universit\"at Rostock, D-18051 Rostock, Germany}
\author{G. S.  Agarwal}\affiliation{Department of Physics, Oklahoma State University, Stillwater, OK, USA}

\date{\today}

\begin{abstract}
	We derive a closed photo-counting formula, including noise counts and a finite quantum efficiency, for photon number resolving detectors based on on/off detectors.
	It applies to detection schemes such as array detectors and multiplexing setups.
	The result renders it possible to compare the corresponding measured counting statistics with the true photon number statistics of arbitrary quantum states.
	The photo-counting formula is applied to the discrimination of photon numbers of Fock states, squeezed states, and odd coherent states.
	It is illustrated for coherent states that our formula is indispensable for the correct interpretation of quantum effects observed with such devices.
\end{abstract}

\maketitle

\section{Introduction}
	Quantum optics delivers a manifold of nonclassical features of quantum light.
	These nonclassical properties can be used for various applications in the field of Quantum Information Technology~\cite{QIT}, Quantum Metrology~\cite{Metrology}, or for highly sensitive measurements, see e.g.~\cite{GEO600}.
	For this reason, a reconstruction of at least some properties of the quantum state is essential, cf. e.g.~\cite{VogelBook,MandelWolf,VogelReview}.
	One major aspect of a quantum state is its photon number distribution~\cite{Yamamoto,Raymer}.
	Sub-Poisson photon statistics renders it possible to go beyond the shot noise limit of classical states.
	Even phase sensitive measurements, such as homodyne detection with a weak local oscillator, require the resolution of the photon number distribution~\cite{HDweakLO1,HDweakLO2}.

	However, photon number resolving (PNR) detectors are not always accessible.
	For example, an avalanche photo-diode in the Geiger mode is an on/off detector~\cite{SinglePhoton}.
	It cannot deliver any information beside the absence or presence of photons.
	Even this tiny bit of information can be hidden when taking into account noise and losses, see e.g.~\cite{DetectorTomography2,DetectorTomography1,NoiseLoss}.
	Surprisingly, a subsequent data analysis is possible to obtain a photon statistics with suppressed noise and losses~\cite{InverseNoise1,InverseNoise2}.

\begin{center}
	\begin{figure}
		\includegraphics*[width=6.3cm]{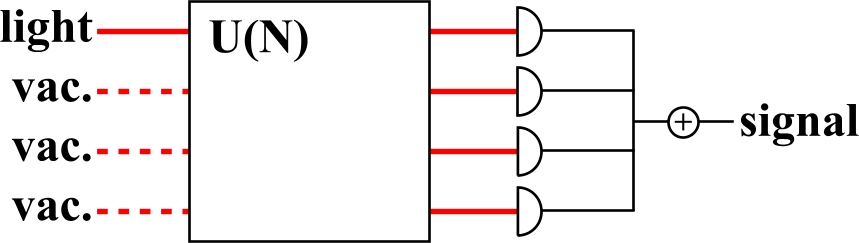}
		\caption{(color online)
		From the left side quantum light enters the measurement scheme.
		The incident light and $N-1$ ports with vacuum input are combined by a $N$-port unitary operation.
		The $N$ output beams are detected via on/off detectors.
		The individual clicks are added up to the total number of counts.
		}\label{Fig:Measure1}
	\end{figure}
	\end{center}

	There are some proposals for PNR detectors based on on/off detectors.
	In Fig.~\ref{Fig:Measure1}, the general idea is given.
	The incoming light is split into weaker signals.
	Each of those weaker signals has only a small probability to include more than one photon.
	Thus, it can be approximately measured by on/off detectors.
	So far the reconstruction of the true photon statistics has been done via iterative maximum-likelihood methods, see e.g.~\cite{Zambra}.
	The masured probabilities can be also directly used to determine, for example, photon number correlations~\cite{Allevi}.

	One experimentally realized method is the use of detector arrays~\cite{Yamamoto,DetArray}.
	Each individual detector in the array is an on/off detector.
	However, simultaneous clicks of some detector deliver additional information about the photon number distribution of the incident quantum state.
	Another experimentally realized PNR detector is the multiplex detection scheme, which is based on a multiple  50:50 splitting of a signal into weaker ones~\cite{TMD1,TMD2,TMD3, TMD5, JumpOnOff, HighCorr}.
	The incoming signal can be split and detected in different spatial or temporal -- time-multiplexing -- modes or bins.

	In this article we will derive a closed analytical counting formula for PNR detectors based on on/off detectors.
	In addition, finite quantum efficiencies and noise count rates will be included in this new formula.
	This closed counting formula can be used to predict the outcome, for example, of a multiplexing or array detector for arbitrary quasimonochromatic quantum states.
	We apply this method to Fock states, squeezed states, and odd coherent states in different realistic scenarios.
	We also discuss the interpretation of data from PNR detectors for determining quantum effects.

	The article is structured as follows.
	In Sec.~\ref{Sec:PCformula}, we derive the counting formula for on/off~detector systems.
	Examples are given in Sec.~\ref{Sec:examples} for perfect and imperfect measurements.
	Systematic error effects of on/off~detector setups are discussed in Sec.~\ref{Sec:error}.
	A summary and conclusions are given in Sec.~\ref{Sec:SC}.

\section{Photo-counting formula}\label{Sec:PCformula}
Any quantum state can be written in terms of the Glauber-Sudarshan~$P$ function as~\cite{GSfunction1,GSfunction2}
\begin{align}\label{Eq:GS}
	\hat\rho=\int d^2\alpha\,P(\alpha)\,|\alpha\rangle\langle\alpha|.
\end{align}
Thus, it is sufficient to consider the effects on pure coherent states only.
The usual photo-counting formula for the true photon statistics reads as
\begin{align}\label{Eq:MandelPCPerfect}
	p_k=\int d^2\alpha\,P(\alpha)\frac{|\alpha|^{2k}}{k!}e^{-|\alpha|^2}=\langle{:}\frac{\hat n^k}{k!}e^{-\hat n}{:}\rangle
\end{align}
where $\hat n$ is the photon number operator, $k$ represents the number of photons, and ${:}\,\cdot\,{:}$ is the normal ordering prescription, cf.~\cite{MandelWolf,VogelBook}.
Taking into account a finite quantum efficiency $\eta<1$ and noise counts $\nu>0$, the photo-counting formula is
\begin{align}\label{Eq:MandelPC}
	p_k=\langle{:}\frac{(\eta\hat n+\nu)^k}{k!}e^{-(\eta\hat n+\nu)}{:}\rangle.
\end{align}
In the following we derive a related closed expression for PNR detectors based on on/off detectors.
As noted above, we restrict our consideration to quasimonochromatic fields, which can be obtained as superpositions of initially monochromatic modes.
The photo-counting formula for broad-band radiation is also given in~\cite{MandelWolf,VogelBook}. It could be used to further generalize the approach given below.

For this purpose let us consider the incident coherent state $|\alpha\rangle$, which is transformed by a unitary operation ${\bf U}(N)$,
\begin{align}
	\hat{\bf a}_{\rm out}={\bf U}(N)\,\hat{\bf a}_{\rm in},
\end{align}
with $\hat{\bf  a}_{\rm in/out}=(\hat a_{1,{\rm in/out}},\dots,\hat a_{N,{\rm in/out}})$ being the annihilation operators for the input/output modes, respectively.
We obtain the input-output relation as
\begin{align}
	|\alpha,0,\dots,0\rangle\mapsto |\psi\rangle=|u_1\alpha,\dots,u_N\alpha\rangle,
\end{align}
with $(u_i)_{i=1}^N$ being the first row of the unitary matrix ${\bf U}(N)$, and therefore $\sum_i |u_i|^2=1$.
In the following we consider the detection of this output light field.
For a single perfect on/off detector, the projection operator valued measure can be given by
\begin{align}
	\hat\pi_0=|0\rangle\langle 0|={:}e^{-\hat n}{:} \quad&\mbox{ and }\quad \hat\pi_1={:}\hat 1-e^{-\hat n}{:},
\end{align}
or equivalently 
\begin{align}\label{Eq:ONOFF}
	\hat\pi_m={:}\left(e^{\hat n}-\hat 1\right)^me^{-\hat n}{:},
\end{align}
for $m=0(1)$ for no(one) click, respectively.
A set of $N$ on/off detectors can be written as
\begin{align}
	\hat\pi_{\bf m}={:}\left(e^{\hat n_1}-\hat 1\right)^{m_1} e^{-\hat n_1}\otimes\dots\otimes\left(e^{\hat n_N}-\hat 1\right)^{m_N} e^{-\hat n_N}{:},
\end{align}
together with the multi-index notion ${\bf m}=(m_1,\dots,m_N)$ and $m_i\in\{0,1\}$.
An output signal of $k$ clicks would mean that $k$ detectors deliver a click, $|{\bf m}|=\sum_i m_i=k$,
\begin{align}
	\hat\Pi_{k}={:}\sum_{|{\bf m}|=k} \bigotimes_{i=1}^{N}e^{-\hat n_i}\left(e^{\hat n_i}-\hat 1\right)^{m_i}{:}\,.
\end{align}

Now, we obtain the measured counting probability for the coherent state as
\begin{align}
	\nonumber p_{k}(\alpha)=&\langle\psi|\hat\Pi_k|\psi\rangle\\
	=&\sum_{|{\bf m}|=k} e^{-\sum_i|u_i\alpha|^2} \prod_{i=1}^N\left(e^{|u_i\alpha|^2}-1\right)^{m_i}.
\end{align}
In addition we may assume that the incident light is split into $N$ modes with uniformly distributed intensities, $|u_i|^2=1/N$.
This assumption yields
\begin{align}
	\nonumber p_{k}(\alpha)=&\sum_{|{\bf m}|=k} e^{-|\alpha|^2} \left(e^{|\alpha|^2/N}- 1\right)^{|{\bf m}|}\\
	=&\frac{N!}{k!(N-k)!}e^{-|\alpha|^2} \left(e^{|\alpha|^2/N}- 1\right)^k.\label{Eq:CountCoh}
\end{align}
For the multiplexing measurement the above assumption is justified due to the $50:50$ beam splitters, which divide the light beam in each step.
This yields that each of the $2^s=N$ bins -- $s$ is the number of divisions -- has the same intensity.
For the detector array, a K\"ohler illumination, as it is known from classical optics, guaranties that the intensity for each of the $d\times d=N$ detectors is almost identical.
In fact, the studied two types of PNR detectors differ only in the practically available number of bins.

Combining Eqs.~(\ref{Eq:GS})~and~(\ref{Eq:CountCoh}), we obtain the counting statistics for arbitrary quantum states in a closed form
\begin{align}
	\nonumber p_k=&\int d^2\alpha\,P(\alpha)\,p_k(\alpha)\\
	=& \langle{:}\frac{N!}{k!(N-k)!}\left(e^{-\frac{\hat n}{N}}\right)^{N-k} \left(\hat1-e^{-\frac{\hat n}{N}}\right)^k{:}\rangle,\label{Eq:ClosedFormula}
\end{align}
where the latter part of the formula expresses the counting statistics in terms of normally ordered expectation values for $k$ counts ($k=0,\dots,N$).
Note that this statistics represents a quantum version of the binomial distribution.

For the inclusion of imperfections, we ristrict our attention to the 
quantum efficiency and noise counts, such as dark counts or scattered light. For the following we assume that 
a given and previously characterized detector is sufficiently well described by these disturbances.
Depending on technical details of the detector, there may occur other imperfections which could be included if desired.

For the presence of losses and noise counts, we derive in a similar way
\begin{align}\label{Eq:Main}
	p_k=\langle{:}\frac{N!}{k!(N-k)!}\left(e^{-(\eta\frac{\hat n}{N}+\nu)}\right)^{N-k} \left(\hat1-e^{-(\eta\frac{\hat n}{N}+\nu)}\right)^k{:}\rangle.
\end{align}
For this purpose we start with a single imperfect on/off detector as $\hat\pi_m={:}\left(e^{\eta \hat n+\nu}-\hat 1\right)^me^{-\eta \hat n+\nu}{:}$ instead of Eq.~(\ref{Eq:ONOFF}).
The formula~(\ref{Eq:Main}) is our main finding, representing the true photo-counting statistics of multiple on/off detectors.
It delivers the probability for $k$ counts from PNR detectors of the type under study.
This expression replaces the photo-counting formula~(\ref{Eq:MandelPC}) for the considered PNR detection scheme.
Note that, our closed analytical expression makes numerical procedures dispensable. So far, the latter were the only available methods for dealing with such counting statistics, cf. e.g. Ref.~\cite{DetectorTomography1,DetectorTomography2}.

\section{Examples}\label{Sec:examples}
In this section, we apply our photo-counting formula.
First, we derive the Fock basis expansion of Eq.~(\ref{Eq:ClosedFormula}).
Using this new expansion, we consider the discrimination of Fock states by $N$ on/off detectors.
We compare the click-counting statistics with the photo-counting statistics without noise and losses for a squeezed state.
Moreover, for an odd coherent state we also deal with the effects of noise and losses.

\subsection{Example without imperfections}
\begin{center}
\begin{figure}[h]
	\includegraphics*[width=4cm]{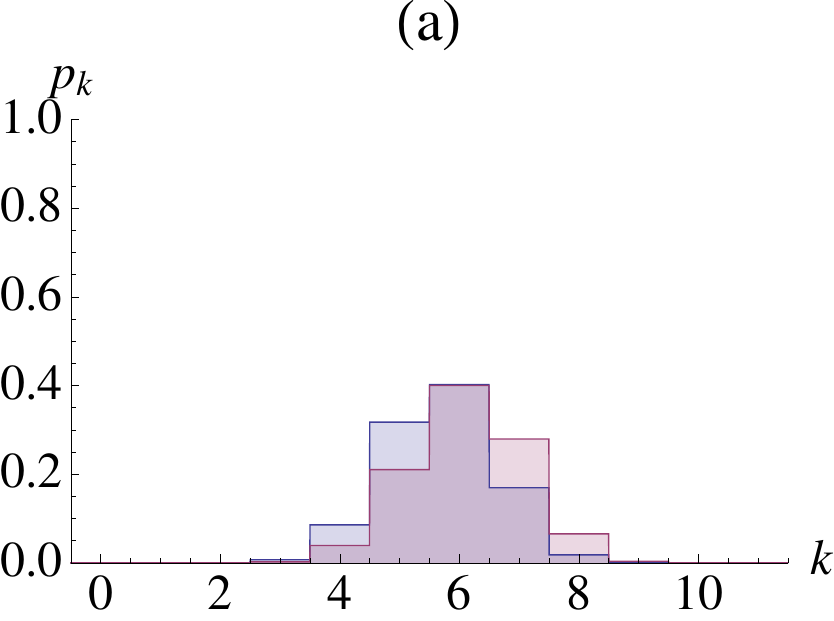}
	\includegraphics*[width=4cm]{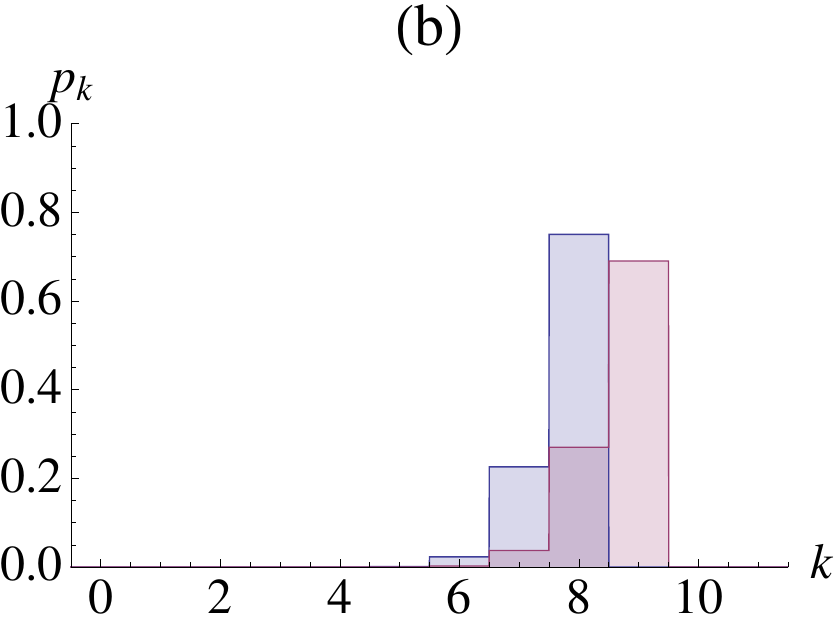}\\[2ex]
	\includegraphics*[width=4cm]{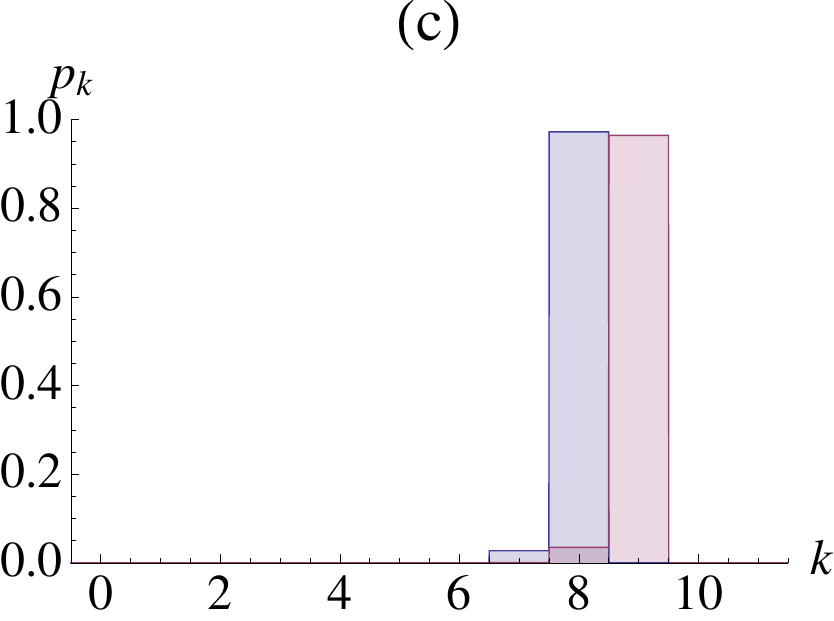}
	\includegraphics*[width=4cm]{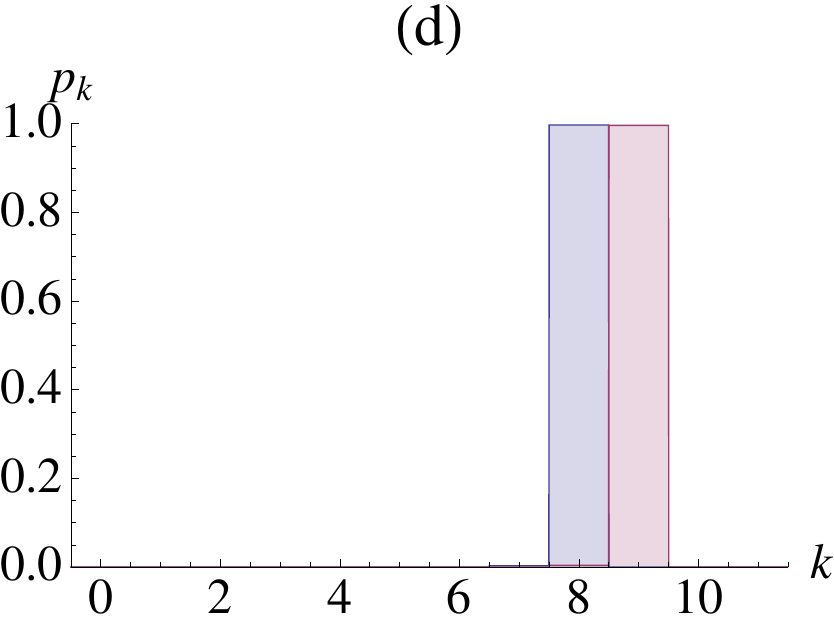}
	\caption{(color online)
	In each plot, we compare the counting statistics of Fock states $|n\rangle$, for $n=8$ (blue) and $n=9$ (red) photons.
	The number of on/off detectors is $N=$10(a),$10^{2}$(b),$10^{3}$(c),$10^{4}$(d).
	}\label{Fig:PhotonTMD}
\end{figure}
\end{center}

First, we consider the discrimination of photon number states $|n\rangle$.
In this paragraph we aim to study effects of the PNR detector itself.
Therefore, we may assume for the time being a quantum efficiency of $\eta=1$ and no noise counts, $\nu=0$.
After some algebra, see Appendix~\ref{App:SomeAlgebra}, it can be seen that the projection operator valued measure for a PNR detector can be written as
\begin{align}
	\hat \Pi_k&=\frac{N!}{k!(N-k)!}{:}\left(e^{\frac{\hat n}{N}}-\hat1\right)^ke^{-{\hat n}}{:}\nonumber\\
	&=\sum_{n=k}^\infty \frac{N!}{k!(N-k)!}\frac{1}{N^{n}}\left(\sum_{j=0}^k\frac{k!(-1)^j(k-j)^n}{j!(k-j)!}\right)|n\rangle\langle n|\nonumber\\
	&=\sum_{n=k}^\infty \left(\frac{N!\left. \partial_x^n\left[e^x-1\right]^k \right|_{x=0}}{N^n k!(N-k)!}\right)
	|n\rangle\langle n|,\label{Eq:14}
\end{align}
using the notion in Eq.~(\ref{Eq:CountCoh}).
In Fig.~\ref{Fig:PhotonTMD}, it can be seen that the number of on/off detectors $N$ must exceed the photon number by a few orders of magnitude in order to sufficiently discriminate different Fock states.

\begin{center}
\begin{figure}[h]
	\includegraphics*[width=4cm]{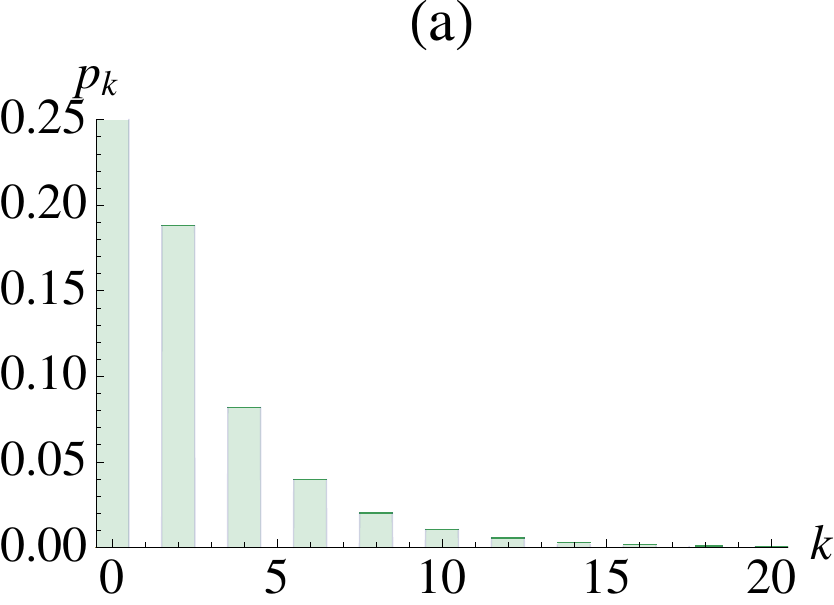}
	\includegraphics*[width=4cm]{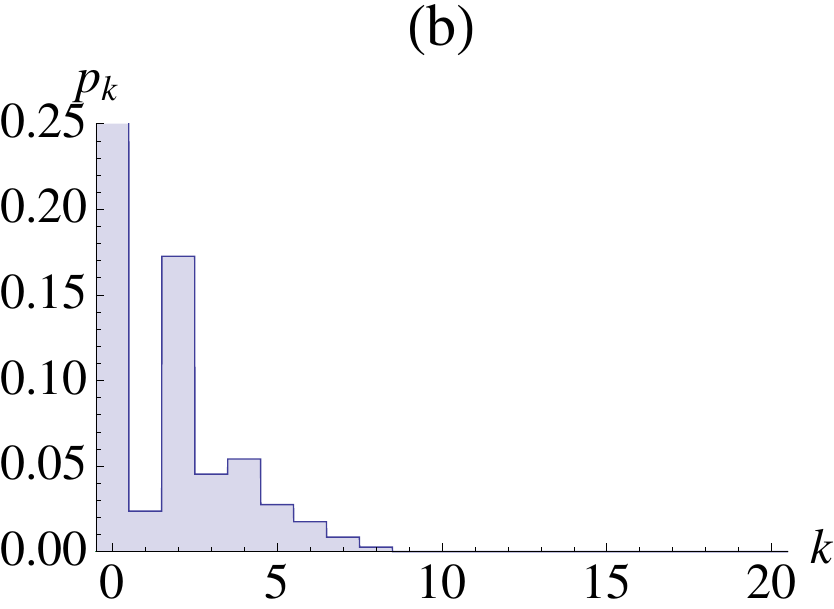}\\[2ex]
	\includegraphics*[width=4cm]{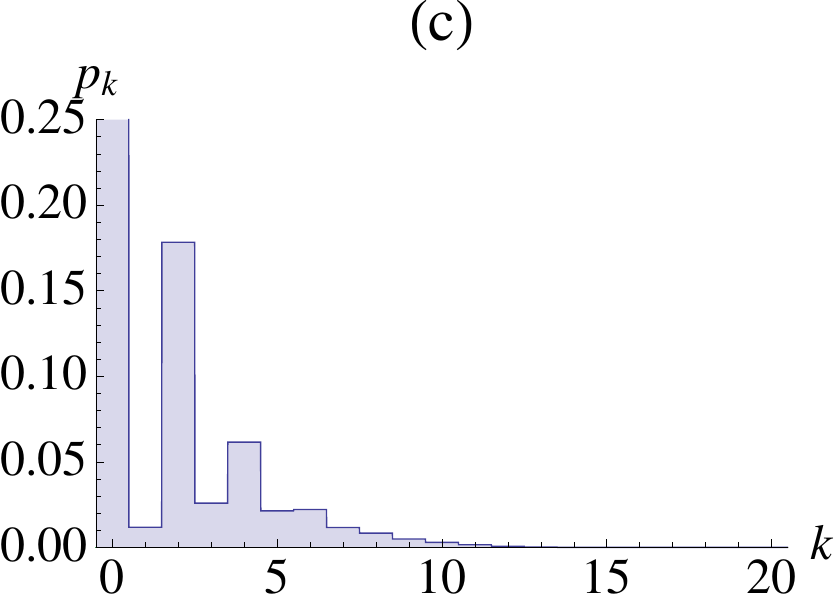}
	\includegraphics*[width=4cm]{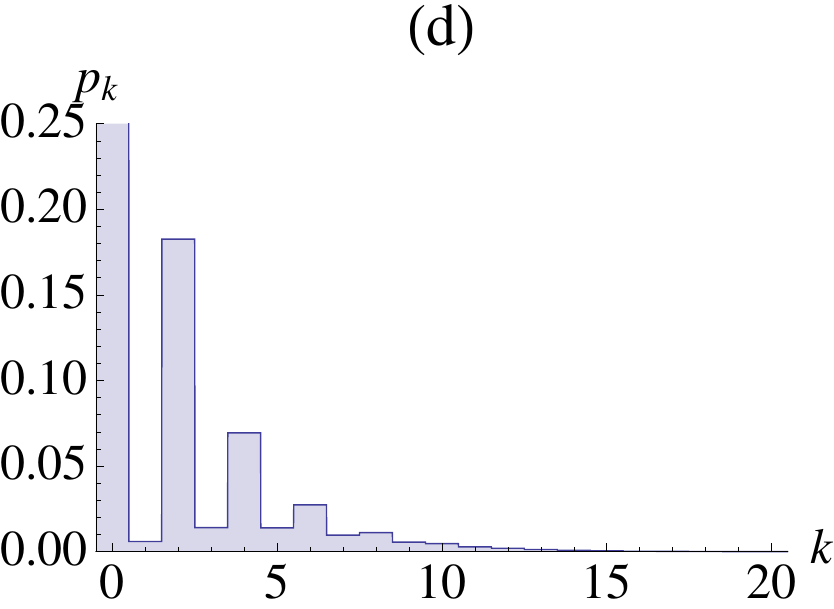}
	\caption{(color online)
	The first figure (a) shows the true photon number statistics (green) of a squeezed state with $\xi=1$.
	The counting statistics obtained by a multiplexing PNR detector (blue) for the same state is given for $s=$3(b),4(c),5(d) multiplexing steps.
	Note that the $p_0$ probability is cut.
}\label{Fig:SqueezedTMD}
\end{figure}
\end{center}

Second, let us consider the counting statistics of a squeezed-vacuum state $|\xi\rangle$~\cite{OsciPhotonNumber1,OsciPhotonNumber2},
\begin{align}
	|\xi\rangle=\frac{1}{\sqrt{\cosh \xi}}\sum_{n=0}^\infty \left(\frac{\tanh \xi}{2}\right)^n\frac{\sqrt{(2n)!}}{n!}|2n\rangle.
\end{align}
This nonclassical quantum state could be the input of a multiplexing measurement with $N=2^s$ bins.
In Fig.~\ref{Fig:SqueezedTMD}, the outcome of a such a measurement for the squeezed-vacuum state is shown.
It can be seen that the probability to have an odd number of clicks is non-zero.
The fact that even photon numbers contribute to odd counting numbers can be understood as a result of the summation of clicks from on/off detectors, cf. Eq.~(\ref{Eq:14}).

\subsection{Example with imperfections}
\begin{center}
\begin{figure}[h]
	\includegraphics*[width=4cm]{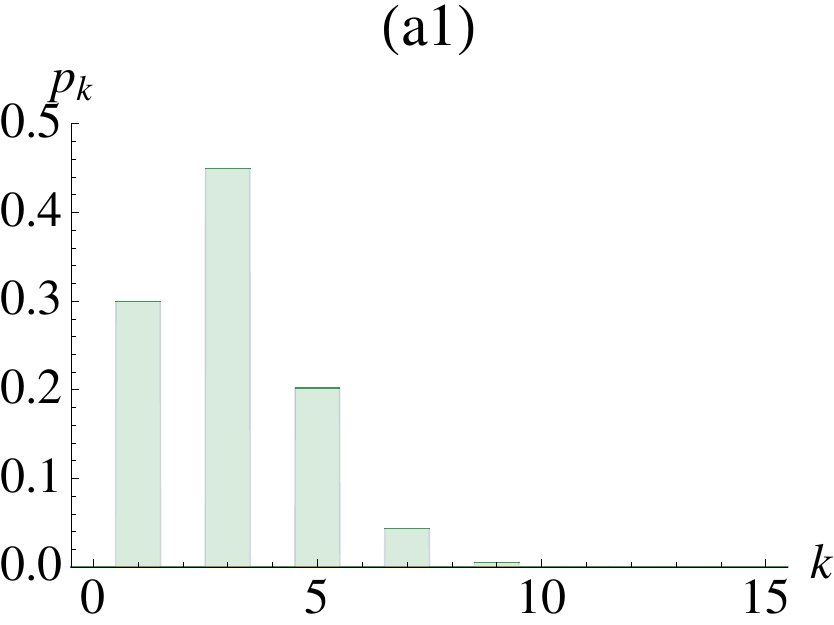}
	\includegraphics*[width=4cm]{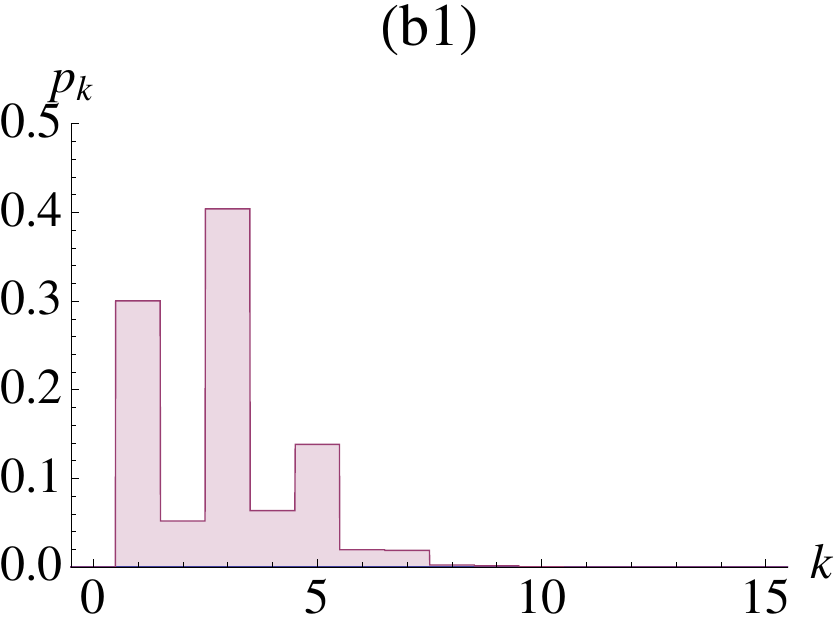}\\[2ex]
	\includegraphics*[width=4cm]{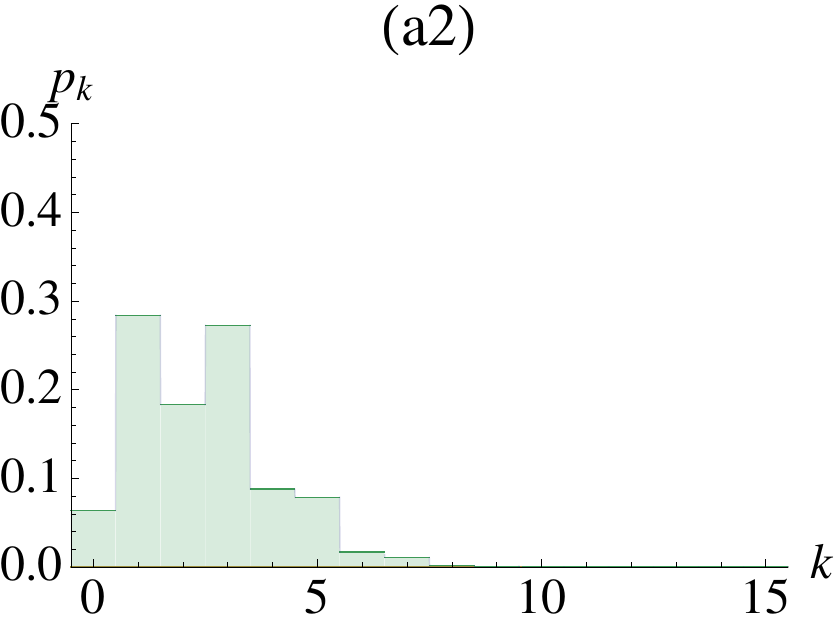}
	\includegraphics*[width=4cm]{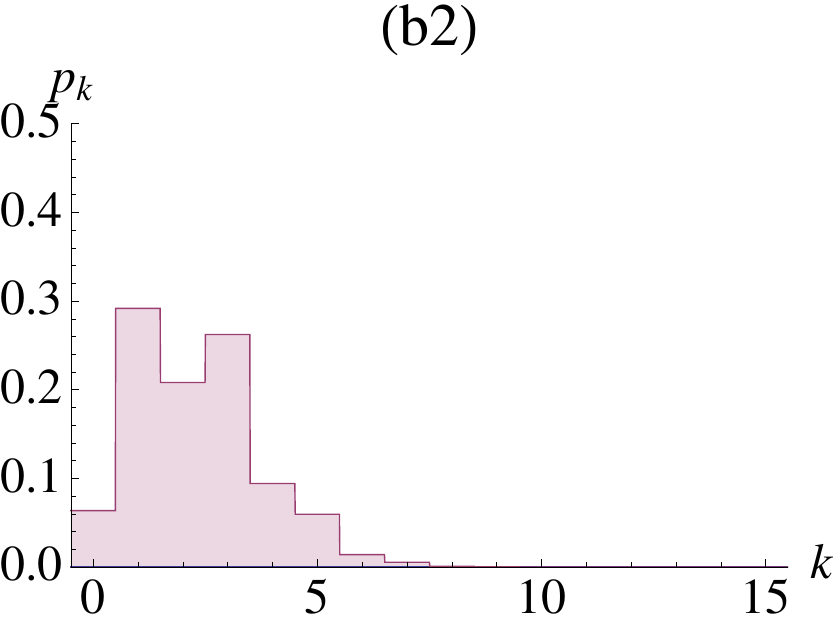}\\[2ex]
	\includegraphics*[width=4cm]{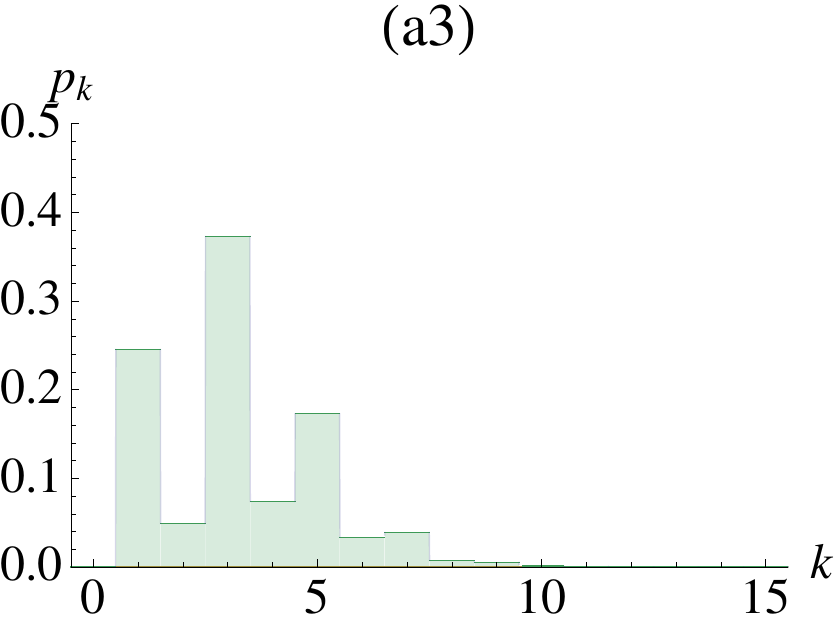}
	\includegraphics*[width=4cm]{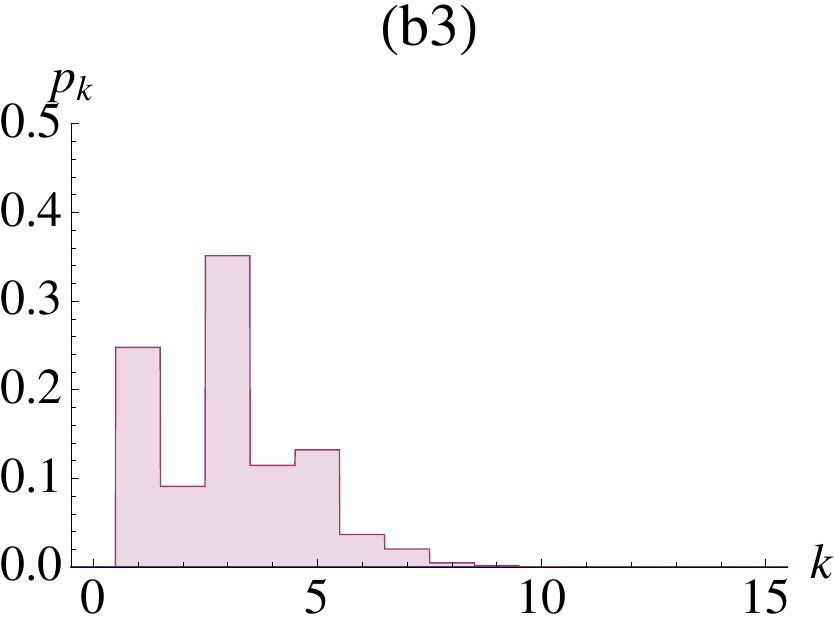}\\[2ex]
	\includegraphics*[width=4cm]{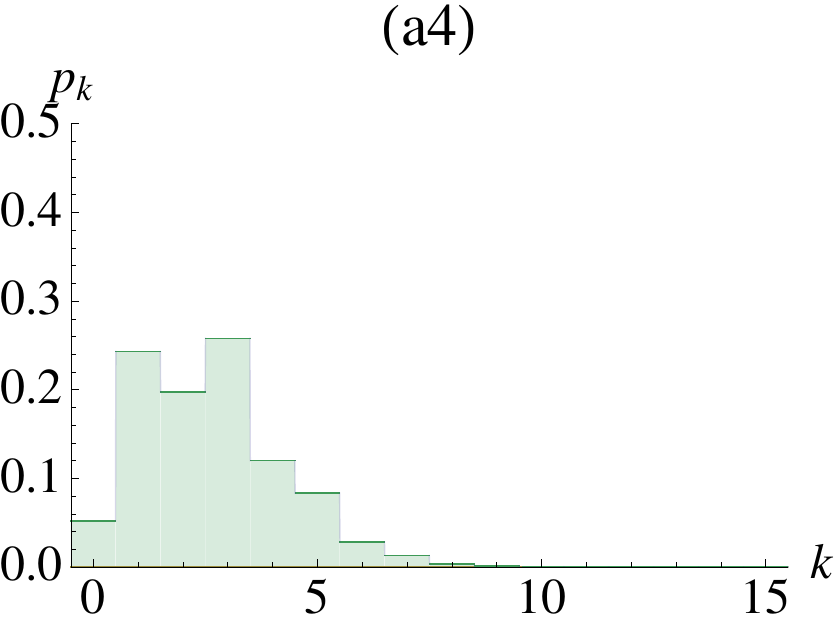}
	\includegraphics*[width=4cm]{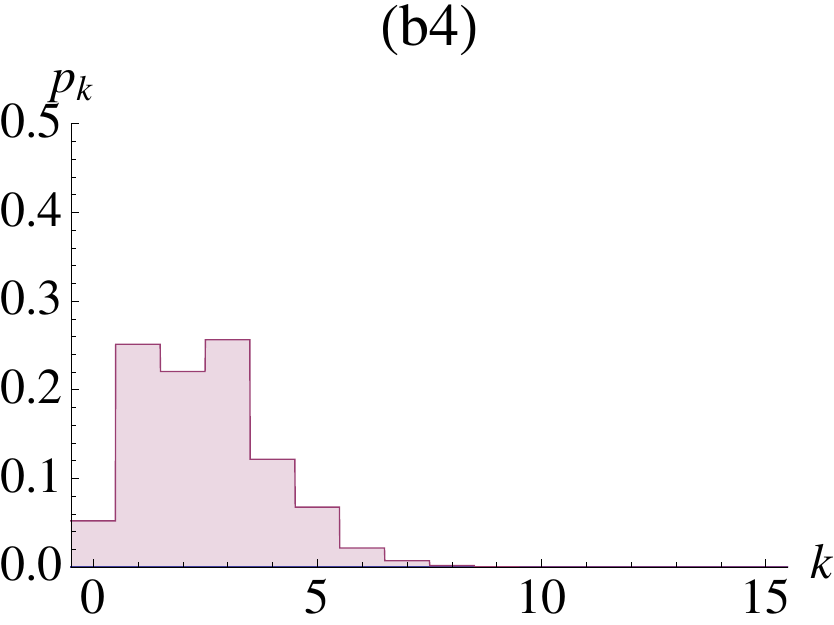}
	\caption{(color online)
	The plots (a1--a4) show the photon statistics (green) of an odd coherent state.
	In comparison, the plots (b1--b4) show the counting statistiscs (red) of PNR detectors based on $N=25$ on/off detectors.
	In the first row -- figures (a1) and (b1) -- we consider a perfect detection.
	In the second row, we include only losses $\eta=0.8$ for both detectors (a2) and (b2).
	In the third row, we consider only noise. For having a comparable result, we considered for (a3) a noise count $\nu=0.2$ and for (b3) $\nu=0.2/N=0.008$.
	The figures (a4) and (b4) include both kinds of the considered imperfections.
}\label{Fig:odd}
\end{figure}
\end{center}

As an example for the effects of noise and losses, we consider the so-called odd coherent state~\cite{Dodonov}
\begin{align}
	|\alpha_{-}\rangle=\mathcal N\left(|\alpha\rangle-|-\alpha\rangle\right),
\end{align}
with a normalization $\mathcal N=\left(2(1-\exp\left[-2|\alpha|^2\right]\right)^{-1/2}$.
The Fock basis expansion of this state only contains odd photon numbers, which is opposite to the case of the squeezed state.
In Fig.~\ref{Fig:odd}, we compare the true photon statistics based on the Mandel formula~(\ref{Eq:MandelPC}) with the counting statistics from PNR detectors based on on/off~detectors, cf. Eq.~(\ref{Eq:Main}).

It can be seen that the perfect detection with $N$ on/off~detectors delivers a non-zero probability for measuring an even click number in the present case, cf. the first row of Fig.~\ref{Fig:odd}.
In this context we have to exclude the case of zero clicks, which has still a zero probability. This can be easily understood on the basis of Eq.~(\ref{Eq:14}), which yields $\hat \Pi_0=|0\rangle\langle 0|$ for $k=0$.
That is, no clicks can only arise from no photons.

We have also considered the case of imperfections in Fig.~\ref{Fig:odd}.
In the second row we only deal with losses due to an imperfect detection efficiency $\eta$.
We see that the change of both statistics is not much different.
This can be basically explained by the idea that losses annihilate some photons, which in particular affects the statistics for small number of counts.
Hence it makes not a big difference, wether the detector is sensitive to a finite number of $k=0,\dots, N$ clicks, compared with an infinite number of $k=0,1,\dots$ photo counts.

In the third row we deal with the case of noise effects only.
Here, we find that the change from the perfect to the noise case is also qualitatively similar in both cases.
For sufficiently weak stray light, the noise effects essentially contribute to the statistiscs at small numbers.
Here we have to note that a fair comparison of both statistics requires that the noise counts of each on/off~detector are only $\nu_{\rm on/off}=\nu/N$.
Since we have to add up the signal of the PNR detector system over the $N$ detectors, 
the total number of noise counts adds up $\nu$.

In the fourths row we combine both effects of imperfections.
Also the combined effects of losses and noise counts are qualitatively similar in the two scenarios.
In practice, the experimenters try to keep dark and stray-light counts as small as possible, so they also contribute mainly to the events at a small number of counts.
Hence the cutoff at a finite but sufficiently large number of on/off detectors is not very important for the effects of the considered imperfections.
From these arguments it becomes clear that the main differences of both detection scenarios are expected to occur due to the cutoff relative to the signal field to be detected, rather than due to the considered imperfections. 

\section{Systematic error effects}\label{Sec:error}
\begin{center}
\begin{figure}[ht]
	\includegraphics*[width=6cm]{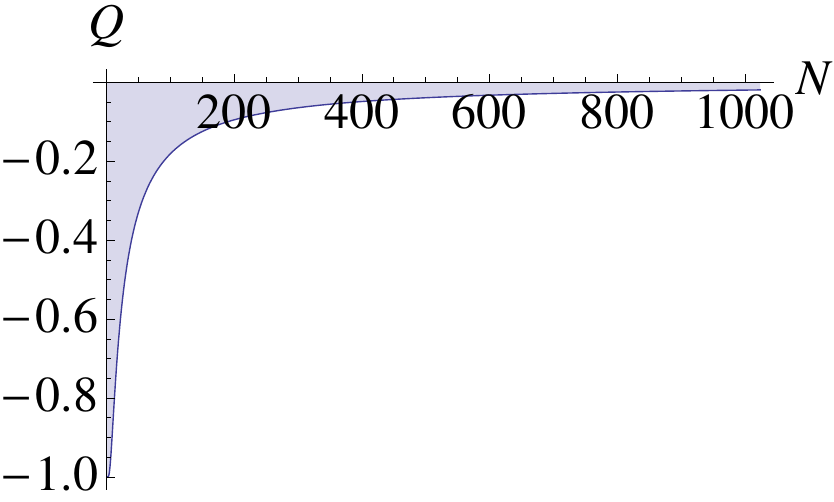}
	\caption{(color online)
	The plot shows the Mandel $Q$ parameter for the counting statistics measured by a PNR detector with $N$ on/off detectors.
	The computed example is a classical coherent state with $|\alpha|^2=20$.
	Note that $N=1$ denotes a measurement with a single on/off detector, and its Q value is close to -1.
}\label{Fig:QMandel}
\end{figure}
\end{center}

Let us address the question of systematical errors from PNR detectors.
For example, the true photon statistics of a quantum state can be used to determine nonclassical effects.
A boundary for classical states can be formulated,
\begin{align}
	Q=\frac{\langle (\Delta\hat n)^2\rangle}{\langle \hat n\rangle}-1\geq0,
\end{align}
which is known as the Mandel $Q$~parameter~\cite{QMandel}.
The violation of this classical condition, $Q<0$, refers to as sub-Poisson photon statistics.
However, determining the $Q$ value from PNR counting statistics delivers misleading results, see Fig.~\ref{Fig:QMandel}.
The plotted range is given for a maximal value of $N=1024$, which corresponds to a multiplexing detector with $s=10$, or a detector array with $d=32$.
The true photon number distribution of the considered coherent state is a Poisson statistics, with $Q=0$.
The plotted negativities are a result of the on/off detectors, and they must not be interpreted as nonclassical effects.
Note that such effects from on/off detectors also appear in measurements of Bell parameters~\cite{FakeOnOff}.
The $Q$ parameter for a coherent state with a mean photon number $|\alpha|^2$ measured by $N$ on/off detectors is $Q=\exp(-|\alpha|^2/N)-1$.
The PNR $Q$ parameter tends to the expected value zero for $N\to\infty$, which is equivalent to the fact that the binomial distribution converges to the Poisson one, see~Appendix~\ref{App:Convergence}.
From Fig.~\ref{Fig:QMandel} it follows that a number of $N=200$ bins is not sufficient to infer Poisson photon statistics, $Q\approx-0.1$, from the measured PNR statistics for $|\alpha|^2=20$.
Remarkably, in this case the number of on/off detectors exceeds the mean photon number of the state by one order of magnitude.
A careful interpretation of quantum effects requires that such systematic errors have to be taken into account, when measuring with PNR detectors.
The photo-counting formula for PNR detectors reveals these systematical errors.

\section{Summary and Conclusions}\label{Sec:SC}
We derived a closed photo-counting formula for photon number resolving detectors based on on/off detectors.
The cases of perfect and imperfect detection have been considered, the latter include finite quantum efficiencies and noise counts.
It has been shown that the resulting photo-counting formula can be interpreted as a quantum version of the binomial distribution.
For large numbers of on/off~detectors, this distribution converges to the well-known photo-counting formula by Mandel.

The derived photo-counting formula has been applied to photon number states, squeezed states, and odd coherent states.
In particular, we considered the discrimination of Fock states in dependence on the number of on/off detectors.
The true photon statistics of a squeezed-vacuum state is compared with that from a photon number resolving detector.
The photo-counting statistics in such a case conceals some properties of the photon number distribution of the squeezed-vacuum state.
The effects of noise and losses for on/off~detection setups in relation to the Mandel formula have been discussed for an odd coherent state.

For the coherent state, we studied the behavior of the Mandel $Q$ parameter and its possible interpretations.
We showed that the resulting counting statistics may infer sub-Poisson light, which is caused by the structure of the detector system itself.
Such systematic errors can be well understood on the basis of our counting formula.

\section*{Acknowledgment}
The authors are grateful to A. A. Semenov for valuable comments and discussions.
This work was supported by the Deutsche Forschungsgemeinschaft through SFB 652.
J. Sperling gratefully acknowledge financial support by the Oklahoma State University.

\appendix
\begin{widetext}
	\section{Computation of POVM elements in Fock basis}\label{App:SomeAlgebra}
	We aim to derive the Fock expansion for the projection operator valued measure of PNR detectors as given in the article, Eq.~(\ref{Eq:14}), by starting with
	\begin{align}
		\hat \Pi_k=&\frac{N!}{k!(N-k)!}{:}\left(e^{\hat n/N}-\hat 1\right)^ke^{-\hat n}{:}\,{;}
	\intertext{taking into account that $|m\rangle\langle m|={:}\frac{\hat n^m}{m!}e^{-\hat n}{:}$, we may write}
		=&\frac{N!}{k!(N-k)!}\sum_{m_1,\dots,m_k\geq1}{:}\frac{\hat n^{\left(\sum_i m_i\right)}}{N^{\left(\sum_i m_i\right)}m_1!\dots m_k!}e^{-\hat n}{:}\\
		=&\frac{N!}{k!(N-k)!}\sum_{m_1,\dots,m_k\geq1}\frac{\left(\sum_i m_i\right)!}{N^{\left(\sum_i m_i\right)}m_1!\dots m_k!}|\left(\sum_i m_i\right)\rangle\langle\left(\sum_i m_i\right)|\\
		=&\frac{N!}{k!(N-k)!}\sum_{n=k}^\infty\frac{1}{N^n}\left[\sum_{m_1,\dots,m_k\geq1;\,|{\bf m}|=n}\frac{n!}{ m_1!\dots m_k!}\right]|n\rangle\langle n|\\
		=&\frac{N!}{k!(N-k)!}\sum_{n=k}^\infty\frac{1}{N^n}\left[\sum_{l_1,\dots,l_k\geq0;\,|{\bf l}|=n-k}\frac{n!}{ l_1!\dots l_k!}\frac{1}{(l_1+1)\dots(l_k+1)}\right]|n\rangle\langle n|;
	\intertext{by using $1/(l+1)=\int_0^1d\tau\,\tau^l$, we obtain}
		=&\frac{N!}{k!(N-k)!}\sum_{n=k}^\infty\frac{n!}{N^n(n-k)!}\left[\int_0^1d\tau_1 \dots\int_0^1d\tau_k\sum_{l_1,\dots,l_k\geq0;\,|{\bf l}|=n-k}\frac{(n-k)!}{ l_1!\dots l_k!}\tau_1^{l_1}\dots\tau_k^{l_k}\right]|n\rangle\langle n|\\
		=&\frac{N!}{k!(N-k)!}\sum_{n=k}^\infty\frac{n!}{N^n(n-k)!}\left[\int_0^1d\tau_1 \dots\int_0^1d\tau_k\left(\sum_{i=1}^k \tau_i\right)^{n-k} \right]|n\rangle\langle n|;
	\intertext{the integration via an induction, $k-1\rightarrow k$, delivers}
		=&\frac{N!}{k!(N-k)!}\sum_{n=k}^\infty\frac{1}{N^n}\left[\sum_{j=0}^k \frac{k!(-1)^j(k-j)^n}{j!(k-j)!}\right]|n\rangle\langle n|\\
		=&\frac{N!}{k!(N-k)!}\sum_{n=k}^\infty\frac{1}{N^n}\left[\sum_{j=0}^k \frac{k!(-1)^j\left.\partial_x^n\left(e^x\right)^{k-j}\right|_{x=0}}{j!(k-j)!}\right]|n\rangle\langle n|\\
		p_k=&\frac{N!}{k!(N-k)!}\sum_{n=k}^\infty\frac{1}{N^n}\left[\left.\partial_x^n\left(e^x-1\right)^k\right|_{x=0}\right]|n\rangle\langle n|.
	\end{align}

	\section{Limit for $N\to\infty$}\label{App:Convergence}
	Let us consider the limit for the photo counting formula of a PNR detector.
	As it is known from classical probability theory, the binominial distribution converges to the Poisson one.
	Thus, the following proof for the quantum analogue is almost identical to its classical version.
	We obtain from the Taylor expansion of Eq.~(\ref{Eq:Main}):
	\begin{align}
		\nonumber p_k=&\langle{:}\frac{N!}{k!(N-k)!}\left(e^{\frac{\hat n}{N}}-\hat 1\right)^{k}e^{-\hat n}{:}\rangle
		=\langle{:}\frac{N!}{k!(N-k)!}\left(\frac{\hat n}{N} \right)^{k}e^{-\hat n}{:}\rangle+\mathcal O\left(\frac{1}{N}\right)\\
		=&\frac{N}{N}\frac{N-1}{N}\dots\frac{N-k+1}{N}\langle{:}\frac{\hat n^k}{k!}e^{-\hat n}{:}\rangle+\mathcal O\left(\frac{1}{N}\right).
	\end{align}
	This yields
	\begin{align}
		p_k\stackrel{N\to\infty}{\rightarrow}\langle{:}\frac{\hat n^k}{k!}e^{-\hat n}{:}\rangle.
	\end{align}
\end{widetext}

\end{document}